%-------------------------------------------------------------------
% Spiked Harmonic Oscillator Potentials
%		H=d^2/dx^2+Bx^2+lambda/x^alpha
% Joint paper Hall-Saad-Keviczky
% spike.tex  19 January 2001 / 5 Jan 2001
% spike.tex  30 August 2001/ 18 Sept final version
% -------------------------------------------------------------------
\def\ptitle{\tiny Spiked harmonic oscillators $\dots$}
% -------------------------------------------------------------------
%  generic unix 12 fonts (lower case names) with no magstep
% --------------------------------------------------------------------
\font\tr=cmr12                          % Our default
\font\bf=cmbx12                         % Redefinition
                         % Redefinition
\font\it=cmti12                         % Redefinition
\font\trbig=cmbx12 scaled 1500          % Main Title
                          % Theorems                       
\font\tiny=cmr10                        % Running title
% --------------------------------------------------------------------
\output={\shipout\vbox{\makeheadline
                                      \ifnum\the\pageno>1 {\hrule}  \fi 
                                      {\pagebody}   
                                      \makefootline}
                   \advancepageno}

\headline{\noindent {\ifnum\the\pageno>1 
                                   {\tiny \ptitle\hfil
page~\the\pageno}\fi}}
\footline{}
% ---------------------------------------------------------------------

\tr 
%--------------------------------------------------------------------
\def\bra{{\rm <}}    % bra ket:  math mode (to replace angle)
\def\ket{{\rm >}}    %   ket  >
  % new line after displayed equations
\def\ni{\noindent}             % noindent
\def\np{\hfil\vfil\break}
\def\htab#1#2{{\hskip #1 in #2}}
\def\bra{{\rm <}} % bra < math mode
\def\ket{{\rm >}} % ket > math mode
\def\hi#1#2{$#1$\kern -2pt-#2} % hyphen \hi{N}{body} = N-body
\def\hy#1#2{#1-\kern -2pt$#2$} % hyphen hy{large}{N} = large-N

%--------------------------------------------------------------------
% SPACING
% -------------------------------------------------------------------
\baselineskip 20 true pt  % draft 15 
\parskip=0pt plus 5pt 
\parindent 0.25in
\hsize 6.0 true in 
\hoffset 0.25 true in 
% 6 in width with 1.25 in margins default = (6.5, 0)
\emergencystretch=0.6 in                 % TEXBook p 107 : allows h-space 
\vfuzz 0.4 in                            % page-length flexibility
\hfuzz  0.4 in                           % line-length flexibility
\vglue 0.1true in
\mathsurround=2pt                        % Default is 2pt
\topskip=24pt                            % Default is 10pt
% ---------------------------------------------------------------------
%  References
% ---------------------------------------------------------------------
\newcount\zz  \zz=0  % switch for printing references
\newcount\q   %  reference number
\newcount\qq    \qq=0  % starting reference number-1   (usually zero)

\def\pref#1#2#3#4#5{\frenchspacing \global \advance \q by 1     % paper reference
    \edef#1{\the\q}{\ifnum \zz=1{\item{$^{\the\q}$}{#2}{\bf #3},{ #4.}{~#5}\medskip} \fi}}

\def\bref #1#2#3#4#5{\frenchspacing \global \advance \q by 1     % book reference
    \edef#1{\the\q}
    {\ifnum \zz=1 { %
       \item{$^{\the\q}$} 
       {#2}, {\it #3} {(#4).}{~#5}\medskip} \fi}}

\def\gref #1#2{\frenchspacing \global \advance \q by 1  % general reference
    \edef#1{\the\q}
    {\ifnum \zz=1 { %
       \item{$^{\the\q}$} 
       {#2.}\medskip} \fi}}

 \def\sref #1{#1}

\def\references#1{\zz=#1
   \parskip=2pt plus 1pt   % default is 0pt plus 1pt       
   {\ifnum \zz=1 {\noindent \bf References \medskip} \fi} \q=\qq
%--------------------------------------------------------------------
\pref{\harr}{E. M. Harrell, Ann. Phys. }{105}{ 379 (1977)}{}
%-----------------------------------------------------------------------------
\pref{\hala}{R. Hall, N. Saad and A. von Keviczky, J. Math. Phys.}{39}{6345-51 (1998)}{}
%-----------------------------------------------------------------------------
\pref{\halb}{R. Hall and N. Saad, J. Phys. A: Math. Gen.}{33}{569 (2000)}{}
%-----------------------------------------------------------------------------
\pref{\halc}{R. Hall and N. Saad, J. Phys. A: Math. Gen.}{33}{5531 (2000)}{}
%-----------------------------------------------------------------------------
\pref{\hald}{R. Hall, N. Saad and A. von Keviczky, J. Phys. A: Math. Gen.}{34}{1169 (2001)}{}
%-----------------------------------------------------------------------------
\pref{\agua}{V. C. Aguilera-Navarro, G.A. Est\'evez, and R. Guardiola, J. Math. Phys.}{31}{99 (1990)}{}
%-----------------------------------------------------------------------------
\pref{\agub}{V. C. Aguilera-Navarro and R. Guardiola, J. Math. Phys.}{32}{2135 (1991)}{}
%-----------------------------------------------------------------------------
\pref{\klaa}{J. R. Klauder, Acta Phys. Austriaca Suppl.}{11}{341 (1973)}{}
%-----------------------------------------------------------------------------
\pref{\Klab}{J. R. Klauder, Phys. lett. B}{47}{523 (1973)}{}
%-----------------------------------------------------------------------------
\pref{\klac}{J. R. Klauder, Science}{199}{735 (1978)}{}
%-----------------------------------------------------------------------------
\pref{\eks}{H. Ezawa, J. R. Klauder, and L. A. Shepp, J. Math. Phys.}{16}{783 (1975)}{}
%-----------------------------------------------------------------------------
\pref{\sim}{B. Simon, J. Functional Analysis}{14}{295 (1973)}{} 
%-----------------------------------------------------------------------------
\pref{\deh}{B. DeFacio and C. L. Hammer J. Math. Phys.}{15}{1071 (1974)}{}
%-----------------------------------------------------------------------------
\pref{\detw}{L. C. Detwiler and J. R. Klauder, Phys. Rev. D}{11}{1436 (1975)}{}
%-----------------------------------------------------------------------------
\pref{\rms}{R. M. Spector, J. Math. Phys.}{8}{2357 (1967)}{}
%-----------------------------------------------------------------------------
\pref{\mdel}{M. de Llano, Rev. Mex. Fis.}{27}{243 (1981)}{}
%-----------------------------------------------------------------------------
\pref{\aguc}{V. C. Aguilera-Navarro, F. M. Fern\'andez, R. Guardiola and J. Ros, J.Phys. A: Math. Gen}{25}{6379 (1992)}{}

\pref{\agud}{V. C. Aguilera-Navarro, A. L. Coelho and Nazakat Ullah, Phys. Rev. A}{49} {1477 (1994)}{}

\pref{\zno1}{M. Znojil, J. Phys. A: Math. Gen.}{15}{2111 (1982)}{}

\pref{\zno2}{M. Znojil, J. Phys. lett.}{101A}{66 (1984)}{}

\pref{\zno3}{M. Znojil, Phys. lett.}{158A}{436 (1991)}{}
\pref{\zno4}{M. Znojil and P. G. L. Leach, J. Math. Phys.}{33(8)}{2785 (1992)}{}

\pref{\sol}{Solano-Torres, G. A. Est\'eves, F. M. Fern\'andez,
and G. C. Groenenboom, J. Phys. A: Math. Gen.}{25}{3427 (1992)}{}

\pref{\fer}{F. M. Fermendez, Phys. Lett. A}{160}{ 51 (1991)}{}

\pref{\flyn}{M. F. Flynn, R. Guardiola, and M. Znojil, Czech. J.
Phys.}{41}{1019 (1993)}{}

\pref{\nag}{N. Nag and R. Roychoudhury, Czech. J. Phys.}{46}{343 (1996)}{} 

\pref{\znoc}{M. Znojil and R. Roychoudhury, Czech. J. Phys.}{48}{1 (1998)}{}

\pref{\esta}{E. S. Est\'evez-Bret\'on and G. A. Est\'evez-Bret\'on,  J. Math. Phys.}{34}{437 (1993)}{}

\pref{\hnas}{R. Hall and N. Saad, Can. J. Phys.} {73}{493
(1995)}{}

\pref{\kila}{J. Killingbeck, J. Phys. A: Math. Gen.}{13}{49 (1980)}{}

\pref{\kilb}{J. Killingbeck, J. Phys. A: Math. Gen.}{13}{L231 (1980)}{}

\pref{\kilc}{J. Killingbeck, J. Phys. B: Mol. Phys.} {15} {829 (1982)}{}

\pref{\kild}{J. Killingbeck, J. Phys. A: Math. Gen.}{14}{1005 (1981)}{}

\pref{\kile}{J. Killingbeck, G. Jolicard and A. Grosjean, J. Phys. A: Math. Gen.}{34}{L367 (2001)}{}

\pref{\kola}{H. J. Korsch and H. Laurent, J. Phys. B: At. Mol. Phys.}{14}{4213 (1981)}{}

\pref{\znoa}{M. Znojil, J. Math. Phys.}{30}{23 (1989)}{}

\pref{\znob}{M. Znojil, J. Math. Phys.}{31}{108 (1990)}{}

\pref{\znoc}{M. Znojil, J. Math. Phys.}{34}{4914 (1993)}{}

\pref{\mill}{H. G. Miller, J. Math. Phys.}{35}{2229 (1994)}{}

\gref{\muod}{O. Mustafa, M. Odeh, e-print quant-ph/006004}{}

\gref{\muod}{O. Mustafa, M. Odeh, e-print quant-ph/9910027}{}

\gref{\jski}{J. Skibi\'nski, e-print quant-ph/0007059}

\pref{\case}{K. M. Case, Phys. Rev.}{80}{797 (1950)}{}

\pref{\spec}{R. M. Spector, J. Math. Phys.}{5}{1185 (1964)}{}

\bref{\gold}{I. I. Gol'dman and D. V. Krivchenkov}{Problems in Quantum mechanics}{Pergamon, London, 1961}{}

\bref{\land}{L. D. Landau and E. M. Lifshitz}{Quantum mechanics: non-relativistic theory}{Pergamon, London, 1981}{}

\bref{\rina}{F. Riesz and B. Sz.-Nagy}{Functional Analysis}{F. Ungar Publishing Co., New York, 1965}{Sec. 124, pages 329-335, especially theorem on page 335.}

\bref{\jowh}{J. Weidmann}{Lineare Operatoren in Hilbertr\"aumen}{B. G. Teubner, Stuttgart 1976} {page 120, Thm. 5.38}

\bref{\jow1}{J. Weidmann}{Lineare Operatoren in Hilbertr\"aumen}{B. G. Teubner, Stuttgart 1976} {page 190, Thm. 7.23, especially statement (iv) as well as last sentence; page 191, thm. 7.24, especially statement (iv)}

\bref{\rin1}{F. Riesz and B. Sz.-Nagy}{Functional Analysis}{F. Ungar Publishing Co., New York, 1965}{pages 275-276, especially last sentence in paragraph on page 276.}

\bref{\doe}{G. Doetsch}{\it Handbuch der Laplace-Transformation, Band I}{Birkh\"auser Verlag, Basel und Stuttgart, 1971}{Thm. 2, pages 34, 37 and 153.}

\bref{\BS}{H. Behnke and F. Sommer}{\it Theorie der Analytischen Funktionen einer Komplexen Ver\"anderlichen} {Springer-Verlag, Berlin, 1976}{Thm. 28, pages 138-139.}

\bref{\slat}{L. J. Slater}{Generalized Hypergeometric
Functions}{University Press, Cambridge, 1966}{}

\bref{\sla}{L. J. Slater}{Confluent Hypergeometric Functions}{University Press, Cambridge, 1960}{}
}% end of ref list

\references{0}    % Initialization of reference numbers
% ------------------------------------------------------------------ end our ref.tex

% ----------------------------
% Preprint list
% ----------------------------
\htab{3.5}{CUQM-88}

\htab{3.5}{math-ph/0109014}

\htab{3.5}{September 2001}
%-------------------------------------------------------------------
%Title Page 
%-------------------------------------------------------------------
\vskip 0.5 true in
\centerline{\bf\trbig Spiked Harmonic Oscillators}
\medskip
\vskip 0.25 true in
\centerline{Richard L. Hall$^\dagger$, Nasser Saad$^\ddagger$ and Attila B. von Keviczky$^\dagger$}
\bigskip
{\leftskip=0pt plus 1fil
\rightskip=0pt plus 1fil\parfillskip=0pt
\obeylines
$^\dagger$Department of Mathematics and Statistics, Concordia University,
1455 de Maisonneuve Boulevard West, Montr\'eal, 
Qu\'ebec, Canada H3G 1M8.\par}

\medskip
{\leftskip=0pt plus 1fil
\rightskip=0pt plus 1fil\parfillskip=0pt
\obeylines
$^\ddagger$Department of Mathematics and Computer Science,
University of Prince Edward Island, 
550 University Avenue, Charlottetown, 
PEI, Canada C1A 4P3.\par}

\vskip 0.5 true in
%---------------------------------------------------------------------------
% Abstract
%---------------------------------------------------------------------------
\centerline{\bf Abstract}\medskip
A complete variational treatment is provided for a family of 
spiked-harmonic oscillator Hamiltonians $H=-{d^2\over dx^2}+B x^2+{\lambda\over x^\alpha}$\quad ($B>0,\ \lambda> 0$), for arbitrary $\alpha>0$. A compact topological proof is presented that the set $S = \{\psi_{n}\}$ of known exact solutions for $\alpha = 2$ constitutes an orthonormal basis of the Hilbert space $L_2(0,\infty)$. Closed-form expressions are derived for the matrix elements of $H$ with respect to $S$. These analytical results, and the inclusion of a further free parameter, facilitate optimized variational estimation of the eigenvalues of $H$ to high accuracy. 


\vfil\eject

%---------------------------------------------------------------------------
% 1. Introduction
%---------------------------------------------------------------------------
\ni{\bf 1. Introduction}\medskip

A family of quantum Hamiltonians known as spiked harmonic oscillators is given by the general Hamiltonian operator 
$$
H=-{d^2\over dx^2}+ x^2+{\lambda\over x^\alpha},\eqno(1.1)
$$
acting in the Hilbert space $L_2(0,\infty)$. Eigenfunctions $\psi \in L_2(0,\infty)$ of $H$ satisfy the Schr\"odinger equation
$$
-\psi^{\prime\prime} +\{x^2+{\lambda\over x^\alpha}\}\psi = E\psi\quad\hbox{ with }\quad\psi(0) = 0.\eqno(1.2)
$$
The function $\psi$ is an eigenfunction corresponding to the eigenvalue $E$ and the condition $\psi(0) = 0$ is called a {\it Dirichlet boundary condition}. The name of the operator derives from the graphical shape of the full potential$^\sref{\harr}$ $V(x)=x^2+{\lambda\over x^\alpha}$, which shows a pronounced peak near the origin for $\lambda>0$. Further, the Hamiltonian Eq.(1.1) is characterized by means of two parameters$^{\sref{\harr}}$, namely $\lambda$ playing the role of a coupling constant and $\alpha\geq 0$ determining the degree of the singularity of the potential at the origin. Therefore, it has been regarded as a two parameter ($\alpha, \lambda$) problem. Recently however, the present authors$^{\sref{\hala}}$ have studied a more general family of Hamiltonians, known as generalized spiked harmonic oscillators, determined by
$$
H=H_0+{\lambda\over x^\alpha},\ H_0=-{d^2\over dx^2}+ B x^2+{A\over x^2}\quad (B>0, A\geq 0)\eqno(1.3)
$$
as an operator in $L_2(0,\infty)$. They argued$^{\sref{\hala-\hald}}$ that the basis, constructed from the exact solutions of the singular Goldman and Krivchenkov Hamiltonian $H_0$ in Eq.(1.3), forms a more effective  starting point for a perturbative variational treatment of the Hamiltonian (1.1) than the basis of the ordinary harmonic oscillator, as used, for example, in earlier works of Aguilera-Navarro et al$^{\sref{\agua-\agub}}$. The objective of the present article is the demonstration of this contention, as well as the establishment of a single variational method suitable for the {\it complete} set of eigenvalues of the Schr\"odinger equation (1.2). `Suitable' means that, regardless of the values of the parameters $\lambda$ and $\alpha$, the presented variational treatment remains valid for all discrete energy eigenvalues $E_n$, $n=0,1,2,\dots$ and further, is also valid in the \hi{N}{dimensional} case with an  arbitrary angular-momentum quantum number $l$. 

The article is essentially self-contained, providing verification of the most relevant results concerning the generalized spiked harmonic oscillators (1.3).  
The proof we here give for the completeness of the set $S = \{\psi_n\}$ of eigenfunctions of $H_0$
is topological in nature and compact.  However, the principal advances
of the present work over Ref.[2] are twofold.  Firstly, we have
managed now to derive closed-forms for the infinite sums representing
the matrix elements of $H$ with respect to $S;$ this, in turn, has
allowed us to use these matrices to find accurate energy upper bounds
quickly.  Secondly, we have included a free parameter $A$ in the
perturbation (1.3) and consequently a final minimization of the 
matrix eigenvalues over $A$ generates even more accurate eigenvalue approximations, or allows us to determine the eigenvalues to a given accuracy with the
use of matrices having smaller dimension $D.$ In Sec. 2, we give a brief history of the spiked harmonic oscillator problem and indicate its relevance. Thereafter we present in Sec. 3 the Gol'dman and Krivchenkov potential$^{\sref{\hala}}$  as an exactly solvable model of singular type, and prove that the exact solutions of its Schr\"odinger equation form a complete orthonormal basis of the Hilbert space $L_2(0,\infty)$. This topological fact is accomplished by means of using abscissas of plain convergence, absolute convergence  as well as of holomorphy of the Laplace Transform, which abscissas are well discussed in the book of Gustav Doetsch. We further develop in Sec. 4 explicit expressions for the matrix elements of the Hamiltonian (1.3) for arbitrary values of $\alpha$ subjected to the restriction $\gamma > {\alpha\over 2}$ with special attention given to $\alpha=2,4$ and $6$, for which we have the most complete analytic results. Herein $\gamma=1+{1\over 2}\sqrt{1+4A}$ arises out of the energy expressions for the exact solutions of the Gol'dman and Krivchenkov Hamiltonian $H_0$, whereas $\gamma > {\alpha\over 2}$ follows from the necessity of taking inner products of functions of type $f(x) = x^{\gamma - (\alpha + 1)/2}exp(-\beta x^2/2)P(x)$ with one another in the Hilbert space $L_2(0,\infty)$, where $P(x)$ is an arbitrary polynomial. This is followed by our variational treatment in Sec. 5, wherein we compare in detail our method with that of the variational approach of Aguliera-Navarro et al$^{\sref{\agua-\agub}}$. An extension to the $N$-dimensional case with arbitrary angular momentum number $l=0,1,2,\dots $ is provided in Sec. 6. Finally, numerical results with detailed comparisons to pervious work, accompanied by concluding remarks, are given in Sec. 7. The difficulty concerning convergence of the eigenvalues approximations calculated from the truncated matrix, in particular the slowness of convergence of the eigenvalues, is looked at in depth for the case of $\alpha = 4$. It is proved that there is a critical value $\lambda_c=5/4$ such that convergence is slow when $\lambda<\lambda_c$ and rapid when $\lambda > \lambda_c$. \medskip
% ------------------------------------------------ 
\noindent{\bf 2. Background and brief history}\medskip
% ------------------------------------------------

There are several reasons for the interest in the spiked harmonic oscillator Hamiltonian$^{\sref
{\harr-\jski}}$. First, it represents the simplest model of certain realistic interaction potentials in atomic, molecular and nuclear physics, and second, its interesting intrinsic properties from the viewpoint of mathematical physics. The potential $V(x)=x^2+{\lambda\over x^\alpha}$ gives rise to a notably long-ago recognized surprise - i. e. no dominance of either of the two interaction potential $x^2$ and ${\lambda\over x^\alpha}$ takes place. Thus, for all values of $\lambda>0$,  the term ${\lambda\over x^\alpha}$ always adds an infinite repulsive barrier near the origin and on the other hand, one can never neglect the $x^2$ term because its absence demolishes the existence of the ground-state energy$^{\sref{\case-\spec}}$. Consequently$^{\sref{\agub}}$, the potential $x^2 +{\lambda\over x^\alpha}$ is like a wide valley extending to $\infty$. Further, the study for small $\lambda$ reveals the presence of two different behaviours depending on the value of $\alpha$. When $\alpha < 5/2$, the ground state energy has a power series expansion$^{\sref{\agub}}$ in terms of $\lambda$. For $\alpha\geq 5/2$, the regular perturbation theory fails badly for this kind of ({\it supersingular}) potential. Another interesting observation is that after the perturbation $\lambda x^{-\alpha}$ is shut off ($\lambda\rightarrow 0$), permanent and irreversible (vestigial) effects of the interaction remain$^{\sref{\hald}}$. The latter effect was first noted by Klauder$^{\sref{\klaa-\klac}}$ who discussed this remarkable phenomenon in connection with non-renomalizable covariant quantum field theories. In particular, potentials $V$, which are sufficiently singular cannot be smoothly turned off ($\lambda \rightarrow 0$) in the Hamiltonian $H=H_0+\lambda V$ to restore the free Hamiltonian $H_0$. This is now know as Klauder's phenomenon$^{\sref{\eks-\deh}}$. Indeed, an obvious approach to obtain informations about the spectrum of the Hamiltonian Eq.(1.1) is to regard the potential $x^{-\alpha}$ as a perturbation of the well-known harmonic oscillator Hamiltonian. By means of the exact solutions $u_n$ of the harmonic oscillator, one attempts to investigate the behavior of the spectrum in terms of the unperturbed eigenfunctions $u_n$ of $H_0$. Detwiler and Klauder$^{\sref{\detw}}$ realized that the normal perturbation theory already fails badly for the first correction of the ground-state energy $(n=0)$ for $\alpha\geq 3$. Examining the asymptotic behavior of the lowest eigenvalues of $H$ by means of variational arguments, they were able to predict the kind of dependence the ground state energy $E=E(\lambda)$ has on the coupling $\lambda$ for $\lambda$ sufficiently small. 

Due to the failure of the Rayleigh-Schr\"odinger series, a modified perturbation theory was required, especially to ascertain the higher-order expressions in the energy expansion in terms of powers of $\lambda$. Harrell$^{\sref{\harr}}$ modified the Rayleigh-Schr\"odinger series by utilizing the standard (WKB)-approximation technique for the lowest few orders. This proving to be quite successful, he continued to developed a special perturbation theory,  now know as singular perturbation theory, and obtained thereby the first few terms of the perturbed $\lambda$-expansion for different values of $\alpha$. This turned out to be a non-power series expansion and in fact was of exactly the same order as of Detwiler and Klauder$^{\sref{\detw}}$. Its drawback is however, that it is valid only for sufficiently small value of the coupling $\lambda$ - i. e. for $\lambda\ll 1$. 

Since the early work of Detwiler and Klauder and Harrell, spiked harmonic-oscillator 
Hamiltonians Eq.(1.1) have become the subject of intensive study lasting over three decades$^{\sref{\hala-\jski}}$. Aguilera-Navarro et al$^{\sref{\agua}}$ managed to obtain a strong coupling perturbative expansion ($\lambda \gg 2$) for the ground state energy. Shortly after, Aguilera-Navarro and Guardiola$^{\sref{\agub}}$ attempted to find a path leading from the weak-coupling regime due to Harrell$^{\sref{\harr}}$ to the strong coupling regime$^{\sref{\agua}}$. Nonetheless, they failed to give a general constructive method to relate the two perturbative regimes for arbitrary values of the exponent $\alpha$. Special attention was given to certain values of $\alpha$. Aguilera-Navarro et al$^{\sref{\aguc}}$ analysed the Hamiltonian (1.1) for $\alpha=1$ around the three regions $\lambda\rightarrow -\infty,\lambda\rightarrow 0$ and $\lambda\rightarrow\infty$ via Rayleigh-Ritz large-order perturbative expansions. Aguilera-Navarro et al$^{\sref{\agud}}$ studied the particular case of  $\alpha=4$ by using a non-orthonormal set satisfying the correct boundary conditions at $x\rightarrow\infty$ and $x\rightarrow 0$.  Exact and approximate (variational) solutions of the ground state energy of the spiked harmonic oscillator problem have also been reported for particular parameter-values in the perturbation potential. Aside from these results, special methods have been developed to compute the eigenvalues numerically with high precision$^{\sref{\kila-\mill}}$.

Most of these results however, concern themselves with different approximation techniques for the ground-state energy for the problem in one spatial dimension. Recently, the $N$-dimensional case has began to attract the attention of many researchers$^{\sref{\halb-\hald,\muod-\jski}}$. In earlier work$^{\sref{\hala-\hald}}$ we have pointed out the advantages of basing our variational analysis on an exactly soluble model, which has itself a singular potential term$^{\sref{\hala}}$. We examine a family of generalized spiked harmonic-oscillator Hamiltonians Eq.(1.1) in terms of a one-dimensional space variable $x\ (0\leq x <\infty)$ with eigenfunctions satisfying Dirichlet boundary condition as stated in Eq.(1.3), that is to say, with wave functions vanishing at the boundaries. The singular orthonormal basis, consisting of the set of exact solutions of $H_0$, serves as a better starting point for the analysis of the Hamiltonian Eq.(1.1). In this paper we use these exact solutions of $H_0$ to provide systematic variational solutions of the spiked harmonic oscillator Hamiltonians Eq.(1.1). We first discuss the tools we implement herein, namely the Gol'dman and Krivchenkov orthonormal basis.
\medskip
% ------------------------------------------------ 
\noindent{\bf 3. Orthonormality of confluent hypergeometric series and Singular potentials}\medskip
% ------------------------------------------------
The purpose of the section is to develop the necessary tools for the variational approach we shall present later. Specifically, we derive in this section the orthonormal basis most suitable for dealing with the Hamiltonian Eq.(1.3). Although the variational method does not {\it per se} require  the the use of a complete orthonormal basis, one can employ in some situations an orthogonal, or even a non-orthogonal$^{\sref{\agud}}$, set of sufficiently smooth functions to attain very effective results. In such cases, however, the variational method shall only have meaning for the particular set of chosen functions$^{\sref{\agud}}$, and the slightest modification of the original problem shall immediately necessitate a new set of functions or at least some modifications. Since we are interested in a set of basis functions admissible for the entire parameter range of $\alpha$ and $\lambda$ in Eqs.(1.1) and (1.3) for our variational method, we develop our solution by using the solution of an exactly solvable singular potential whose singularity coincides with that of the spiked harmonic oscillator. Therefore, we use an exactly solvable singular Hamiltonian known as the Gol'dman and Krivchenkov Hamiltonian$^{\sref{\gold-\land}}$ 
$$
-\psi^{\prime\prime}+V_0\bigg({a\over x}
-{x\over a}\bigg)^2\psi=E_n\psi\  x\in[0,\infty)\hbox{ with }\psi(0)=0,\eqno(3.1)
$$
namely $\psi$ satisfies the Dirichlet boundary condition. The energy spectrum in terms of the parameters $V_0$ and $a$ is given
by$^{\sref{\gold}}$ 
$$
E_n={4\over a}\sqrt{V_0}\bigg\{n+{1\over 2}+{1\over 
4}\bigg(\sqrt{1+4V_0a^2}-2a\sqrt{V_0}\bigg)\bigg\},
\eqno(3.2)$$
whereas the exact wavefunctions take the form
$$
\psi_n(x)=C_nx^\nu e^{-{1\over 2}{\sqrt{V_0}\over a}x^2}{}_1F_1(-n,\nu+1,{\sqrt{V_0}\over a}x^2)
\eqno(3.3)
$$
with $\nu = {1\over 2}(1+\sqrt{1+4V_0a^2})$. In terms of the 
Pochhammer symbols
$$(a)_0=1,\quad (a)_k=a(a+1)(a+2)\dots(a+k-1)={{\Gamma(a+k)}\over {\Gamma(a)}},\quad 
k=1,2,\dots$$
expressed in terms of the gamma function $\Gamma$, the previous function ${}_1F_1$, known as the confluent hypergeometric function$^{\sref{\sla}}$, is defined by
$${}_1F_1(-n;b;z)=\sum\limits_{k=0}^n {{(-n)_kz^k}\over {(b)_kk!}}.\eqno(3.4)$$
To simplify the notation, we introduce parameters $B=V_0a^{-2}$ and $A=V_0a^2$, 
and obtain thereby an exact solution to the one-dimensional Schr\"odinger equation with the 
singular potential
$$
V(x)=B x^2+{A\over x^2}, \quad B>0, A\geq 0,\eqno(3.5)$$
whose energy spectrum is now given in terms of parameters $A$ and $B$ by
$${E}_n=2\beta(2n+\gamma),\quad n=0,1,2,\dots,\eqno(3.6)$$
wherein $\beta=\sqrt{B}$ and $\gamma=1+{1\over 2}\sqrt{1+4A}$. The wave functions in this case have the form
$$
\psi_n(x)=C_n x^{\gamma-{1\over 2}} e^{-{1\over 
2}\beta x^2}{}_1F_1(-n;\gamma;\beta x^2)\quad\hbox{ for }\quad n=0,1,2,\dots.\eqno(3.7)$$ 
The constant $C_n$ in Eq.(2.7) is determined from the normalization condition
$$\int_0^{\infty}\psi_n^2(x)dx=1.\eqno(3.8)$$
In order to compute this integral, we need the following lemma, which is a generalization of formula f6 in the appendix of the book {\it Quantum Mechanics} by Landau and Lifshitz$^{\sref{\land}}$.

\noindent{\bf Lemma 1:} For $\gamma>{0}$, and $m,n=0,1,2,\dots$ 
$$\int\limits_0^\infty x^{2\gamma-1} e^{-\beta x^2}{}_1F_1(-n;\gamma;\beta x^2)
{}_1F_1(-m;\gamma;\beta x^2)dx={1\over 2}{n! \Gamma(\gamma)\over \beta^\gamma (\gamma)_n}\delta_{mn},
\eqno(3.9)
$$
where $\delta_{mn}=0$ for $m\neq n$ and 1 for $m=n$.

\noindent{Proof:}
We denote the indefinite integral of the right hand side of Eq.(3.9) by $I_{mn}$ and have 
by means of the series representation Eq.(3.4) of the confluent hypergeometric function ${}_1F_1$ that
$$I_{mn}=\sum\limits_{k=0}^m\sum_{l=0}^n {{(-m)_k(-n)_l}\over 
(\gamma)_k(\gamma)_l}{\beta^{k+l}\over k! l!}\int_0^\infty 
x^{2\gamma+2k+2l-1}e^{-\beta x^2}dx.
$$
Further, after resorting to the integral representation of the gamma function
$$\Gamma(x)=\int_0^\infty e^{-t}t^{x-1}dt,\quad x>0,$$ 
and a change of variables, we obtain for $2\gamma+2k+2l-1>0$ that
$$\eqalign{
I_{mn}&={1\over 2\beta^\gamma}\sum\limits_{k=0}^n\sum_{l=0}^m {{(-n)_k(-m)_l}\over 
(\gamma)_k(\gamma)_l}{\Gamma({\gamma+k+l})\over k! l!}\cr
&={1\over 2\beta^\gamma}
\sum\limits_{k=0}^n
\bigg[\sum_{l=0}^m {{(-m)_l(\gamma+k)_l}\over 
(\gamma)_l\ l!}\bigg]
{(-n)_k\Gamma({\gamma+k})\over  (\gamma)_k\ k!}.
}$$
The finite sum inside the bracket is just the series representation of the terminated hypergeometric function ${}_2F_1$, and therefore
$$
I_{mn}={1\over 2\beta^\gamma}
\sum\limits_{k=0}^n
{}_2F_1(-m,\gamma+k;\gamma;1){(-n)_k\Gamma({\gamma+k})\over  (\gamma)_k\ k!}.
$$
Applying Chu-Vandermonde's theorem$^{\sref{\slat}}$ on summing the series ${}_2F_1$ with unit argument, we get
$$
I_{mn}={\Gamma(\gamma)\over 2\beta^\gamma(\gamma)_m}
\sum\limits_{k=0}^n {(-k)_m(-n)_k\over  \ k!},\eqno(3.10)
$$
wherein we have invoked the identity 
$$
(-k)_n=
\cases{
{(-1)^n k!\over (k-n)!}
& if $0\leq n\leq k$,\cr
0
& if $n> k$
.\cr}\eqno(3.11)
$$
On account of Eq.(3.11), the product $(-k)_m(-n)_k$ in Eq.(3.10) differs from zero only for $n=k=m$, and thus 
$\sum\limits_{k=0}^n {(-k)_m(-n)_k\over  \ k!}=n!$ if $n=m$ or 0 if  $n\neq m$, which terminates the proof of our lemma\ \P. 
\medskip
The  reader should realize at this point the connection between the above lemma and the orthonormality relation of Laguerre polynomials. In fact, because of the relation$^{\sref{\sla}}$
$${}_1F_1(-n;\gamma+1;x)={n!\over (\gamma+1)_n}L_n^{(\gamma)}(x),$$
Eq. (3.9) is just another form of the orthonormality condition for Laguerre ploynomials. Thus the normalization constants $C_n$ of Eq.(3.7) can now be determined via Lemma 1 and the normalization condition Eq.(3.8), and this entails
$$
C_n^{-2}={1\over 2}{n! \Gamma(\gamma)\over \beta^\gamma (\gamma)_n},
\eqno(3.12)$$
and the normalized wavefunction Eq.(3.7) now reads as follows
$$
\psi_n(x)=(-1)^n\sqrt{{2\beta^\gamma (\gamma)_n}\over n! \Gamma(\gamma)} x^{\gamma-{1\over 2}} e^{-{1\over 
2}\beta x^2}{}_1F_1(-n;\gamma;\beta x^2).\eqno(3.13)
$$
The alternating coefficients $(-1)^n$ were introduced in the definition of $\psi_n(x)$  to guarantee a smooth transition by means of the identity
$${}_1F_1(-n,{3\over 2},x^2)={(-1)^n\over 2x}{n!\over (2n+1)!}H_{2n+1}(x),\eqno(3.14)$$
to the odd solutions of the harmonic oscillator problem, namely the case $A=0$ in Eq.(3.4). From the immediately preceeding we arrive at

\noindent{\bf Theorem 1:} For $\psi_n(x)$ defined by Eq.(3.13), the following orthonormality relations
$$\int\limits_0^\infty \psi_n(x)\psi_m(x)dx = \delta_{mn},\quad m,n=0,1,2,\dots$$
hold in $L_2(0,\infty)$.
\medskip 
It may seem that the Hamiltonian $H_0$ is self-adjoint, whence we could assume the 
completeness of the normalized eigenfunctions $\{\psi_n(x)\}_{n=0}^\infty$ from the general theory of self-adjoint operators.  However, we are inhibited from doing this because of the domain of definition of $H_0$. The Hamiltonian operator $H_0$ in the Hilbert space $L_2(0,\infty)$ cannot have all of $L_2(0,\infty)$ as its domain of definition on account of the presence of the second derivative  in $H_0$. Consequently, the totality $C_0^\infty(0,\infty)$ of infinitely differentiable complex valued functions on $(0,\infty)$ with compact support is initially assumed to be the domain of definition of $H_0$. $C_0^\infty(0,\infty)$ lies dense in $L_2(0,\infty)$ and $H_0$ is formally adjoint to itself, but not necessarily self-adjoint. To make $H_0$ self-adjoint, we invoke its Friedrichs extension$^{\sref{\rina-\jowh}}$, which is a self-adjoint and exists in principle. If we consider $H_0$ to stand for its Friedrichs extension, then we need to know its spectrum $\sigma(H_0)=\sigma_d(H_0)\cup \sigma_e(H_0)$ - i.e. the disjoint union of its discrete and essential spectrums. $L_2(0,\infty)$ is the direct sum of the eigenspaces of $H_0$ if and only if, the essential spectrum is empty$^{\sref{\jow1-\rin1}}$. We establish $\sigma_e(H_0)=\emptyset$ by demonstrating completeness for the system $\{\psi_n(x)\}_{n=0}^\infty$ in the Hilbert space $L_2(0,\infty)$ of all square integrable functions over the interval $(0,\infty)$ for $H_0$ in its Friedrichs extension form, by means of the well-known Hahn-Banach theorem.
\medskip
\noindent{\bf Theorem 2:} The set of $L_2(0,\infty)$-functions $\{\psi_n(x)\}_{n=0}^\infty$ defined by Eq.(2.13), is a complete orthonormal basis for the Hilbert space $L_2(0,\infty)$.

\noindent{Proof:} The orthonormality of $\{\psi_n(x)\}_{n=0}^\infty$ follows from Theorem(1). To prove completeness, we proceed as follows. On account of the definition ((3.13)) of $\psi_n(x)$ in terms of the hypergeometric function 
$${}_1F_1(-n, \gamma, \beta x^2) = \sum_{k=0}^n {n\choose k}{(-\beta)^{n-k} \Gamma(\beta)\over \Gamma(\gamma+n-k)}x^{2(n-k)}
$$
we can express each of the functions ${}_1F_1(-n, \gamma, \beta x^2)(n \geq 0)$ uniquely as a finite linear combination of $x^{2n}(n \geq 0)$ as well as conversely. This in terms of the span, which we denote by $(L.H.)$, means that $(L.H.)({}_1F_1(-n, \gamma, \beta x^2)(n \geq 0))= (L.H.)(x^{2n}(n \geq 0))$. We multiply each member of these two sets by $x^{\gamma - 1/2}\exp(-\beta x^2/2)$, and thereby obtain a linear subspace of the Hilbert space $L_2(0,\infty)$, whose topological closure satifies
$$\overline{(L.H.)(\cdot^{\gamma - 1/2}{}_1F_1(-n, \gamma, \beta \cdot^2)\exp(-\beta \cdot^2/2)(n \geq 0))} = \overline{(L.H.)(\cdot^{\gamma - 1/2 + 2n}\exp(-\beta \cdot^2/2)(n \geq 0))},
$$  
and further, orthogonal complementation yields
$$[(L.H.)(\cdot^{\gamma - 1/2}{}_1F_1(-n, \gamma, \beta \cdot^2)\exp(-\beta \cdot^2/2)(n \geq 0))]^\bot = [(L.H.)(\cdot^{\gamma - 1/2 + 2n}\exp(-\beta \cdot^2/2)(n \geq 0))]^\bot.
$$
Now we turn to demonstrating that $[(L.H.)(\cdot^{\gamma - 1/2}{}_1F_1(-n, \gamma, \beta \cdot^2)\exp(-\beta \cdot^2/2)(n \geq 0))]^\bot = {0}$, in other words $\{C_n \cdot^{\gamma - 1/2}{}_1F_1(-n, \gamma, \beta \cdot^2)\exp(-\beta \cdot^2/2)\}_{n=0}^\infty$ constitutes an orthonormal basis the Hilbert space $L_2(0,\infty)$ with $C_n$'s given by (3.12). Assuming that $\Phi \in L_2(0,\infty)$ with $\Phi\ \bot\ (L.H.)(\cdot^{\gamma - 1/2}{}_1F_1(-n, \gamma, \beta \cdot^2)\exp(-\beta \cdot^2/2)(n \geq 0))$ leads directly to 

$$< \cdot^{\gamma - 1/2 + 2n}\exp(-\beta \cdot^2/2)\ |\ \Phi > = \int_0^\infty x^{\gamma - 1/2 + 2n}\exp(-\beta x^2/2)\overline{\Phi(x)}dx = 0\ (n \geq 0).
$$
We define 
$$f(s)\equiv \int_0^\infty x^{\gamma - 1/2}\exp(-s x^2/2)\overline{\Phi(x)}dx 
=  
\int_0^\infty e^{-st}F(t)dt,\hbox{ where }  
F(t)\equiv t^{\gamma/2 - 3/4}\overline{\Phi(\sqrt{t})}/2.
$$
Out of the fact that $\Phi \in L_2(0,\infty)$ follows  
$$\eqalign{\int_0^\infty |e^{-st}F(t)|dt &= \int_0^\infty 1/2 t^{\gamma/2 - 3/4}|\Phi(\sqrt{t})|\exp(-(\Re s)t)dt\cr
& = 
\sqrt{2}\int_0^\infty t^{\gamma/2 - 1/2}\exp(-(\Re s)t)|\Phi(\sqrt{t})|{1\over{\sqrt{2}}}t^{-1/4}dt\cr 
&\leq 
\sqrt{2}\sqrt{\int_0^\infty |t^{\gamma/2 - 1/2}\exp(-(\Re s)t)|^2dt}\times \sqrt{\int_0^\infty |\Phi(\sqrt{t})|^2dt} \cr
&= \sqrt{2}\sqrt{\int_0^\infty \exp(-2(\Re s)t)t^{\gamma - 1}dt}\times \sqrt{\int_0^\infty |\Phi(x)|^2dx} \cr
&= \sqrt{2} \sqrt{{1 \over{(2\Re s)^\gamma}}\int_0^\infty e^{-t}t^{\gamma - 1}dt}\times \parallel \Phi \parallel_{L_2(0,\infty)} \cr
&= \sqrt{2(\Re s)^{-\gamma}\Gamma(\gamma)}\parallel \Phi \parallel_{L_2(0,\infty)} < \infty.\cr} $$
Herewith the abscissa $\tilde\alpha$ of absolute convergence$^{\sref{\doe}}$ of the Laplace Integral $f(s)=  {\cal L} \{F\}$, namely the smallest real number ${\tilde\alpha}$, such that $\int_0^\infty |e^{-st}F(t)|dt < \infty$ for all $\Re s > \tilde\alpha$, has the property that $\tilde\alpha \leq 0$. In consequence of our substitution $x=\sqrt{t}$, we have further that
$$
< \cdot^{\gamma - 1/2 + 2n}\exp(-\beta \cdot^2/2)\ |\ \Phi > = (-1)^n \int_0^\infty (-t)^n e^{-(\beta/2)t}F(t)dt = f^{(n)}(\beta/2) = 0 \quad (n \geq 0).\eqno(3.15)
$$
Since $f(s)$ is definitely holomorphic in the half-plane $\Re s > 0$, because the abscissa of holomorphy $\tilde\chi$ satisfies $\tilde\chi \leq \tilde\beta \leq \tilde\alpha$ ($\tilde\beta$ being the abscissa of ordinary convergence of the Laplace Integral$^{\sref{\doe}}$ ${\cal L} \{F\}$), $f(s)$ shall clearly have a Taylor series expansion 
$$
f(s) = \sum_{n=0}^\infty (n!)^{-1}f^{(n)}(\beta/2)(s - \beta/2)^n = 0\  (|s - \beta/2| < \beta/2)\eqno(3.16)
$$
in terms of Eq.(3.15) about the point $\beta/2 > 0$. The radius of convegence of this Taylor series is at least a big as $\beta/2$, since it may happen that the abscissa of holomorphy$^{\sref{\doe}}$ $\tilde\chi$ of the Laplace Integral $f(s) = {\cal L} \{F\}$ is definitely less than the abscissa $\tilde\beta$ of ordinary convergence, not to mention that of absolute convergence $\tilde\alpha$. Thus we have that $f(s) \equiv 0$ in the open disc $|s - \beta/2| < \beta/2$ of the s complex plane. By means of the identity theorem of holomorphic functions$^{\sref{\BS}}$, we have that the Laplace Integral $f(s) = {\cal L} \{F\} \equiv 0$ in the half-plane of holomorphy $\Re s > \tilde\chi$; and more so in the holomorphy domain of $f$, which contains its half-plane of holomorphy. The uniqueness of the Laplace-Transform implies that $F(t) \equiv {1/2}t^{\gamma/2 - 3/4}\Phi(\sqrt{t}) = 0$ a. e. in $t$ on the interval $[0, \infty)$. Hence, $\int_0^\infty |\Phi(\sqrt{t})|^2 ({2\sqrt{t}})^{-1}dt =\int_0^\infty |\Phi(x)|^2dx = 0$, namely $\Phi = 0$ a. e. on $(0,\infty)$, which is to say that $\{C_n \cdot^{\gamma - 1/2}{}_1F_1(-n, \gamma; \beta \cdot^2)\}_{n=0}^\infty$ constitutes a complete orthonormal basis of the Hilbert space $L_2(0,\infty)$\ \P.    

\medskip
% ------------------------------------------------ 
\noindent{\bf 4. The matrix elements for singular potentials}\medskip
% ------------------------------------------------
\noindent It is well-known that the Schr\"odinger equation, even in the one-dimensional case, rarely possesses an exact (analytic) solution. Consequently, a multitude of arduous numerical techniques have been implemented to ascertain its energy eigenvalues over several decades; from among these, we mention matrix diagonalization. Our primary aim in this section is to find the matrix elements of the Hamiltonian Eq.(1.3), whose calculation is achieved by means of 

\noindent{\bf Lemma 2:}  
If $2\gamma-\alpha > 0$, then for all pairs of non-negative integers and $m$ and $n$ we have that 
$$\eqalign{\int\limits_0^\infty x^{2\gamma-\alpha-1} e^{-\beta x^2}{}_1F_1(-n,\gamma,\beta x^2)&
{}_1F_1(-m,\gamma,\beta x^2)dx=\cr{\beta^{{\alpha\over 2}-\gamma}\over 2}{({\alpha\over 2})_n\Gamma(\gamma-{\alpha\over 2})\over (\gamma)_n}
&{}_3F_2(-m,{\gamma-{\alpha\over 2}},{1-{\alpha\over 2}};\gamma,1-{\alpha\over 2}-n;1).\cr}
\eqno(4.1)
$$
\noindent{Proof:} Let $I_{mn}$ denote the infinite integral on the left hand side of Eq.(4.1).
Using the series representation Eq.(3.4) of the confluent hypergeometric function ${}_1F_1$ yields 
$$
I_{mn}=\sum\limits_{k=0}^m\sum_{l=0}^n {{(-m)_k(-n)_l}\over 
(\gamma)_k(\gamma)_l}{\beta^{k+l}\over k! l!}\int_0^\infty 
x^{-\alpha+2\gamma+2k+2l-1}e^{-\beta x^2}dx.\eqno(4.2)
$$
By resorting to the integral representation of gamma function, we obtain under the condition $-{\alpha\over 2}+\gamma+k+l>0$ that
$$\eqalign{I_{mn}&={1\over 2}\sum\limits_{k=0}^m \sum_{l=0}^n {{(-m)_k(-n)_l}\over (\gamma)_k(\gamma)_l}
{\beta^{{\alpha\over 2}-\gamma}\over k! l!}\Gamma(-{\alpha\over 2}+\gamma+k+l)\cr
&={1\over 2} \beta^{{\alpha\over 2}-\gamma}\sum\limits_{k=0}^m
\bigg[\sum_{l=0}^n{{(-n)_l\Gamma(-{\alpha\over 2}+\gamma+k+l)}\over (\gamma)_l\ l!}\bigg]
{(-m)_k\over (\gamma)_k}{1\over k!}.}\eqno(4.3)
$$
This relation we rewrite in terms of the Pochhammer symbols, use $\Gamma(-{\alpha\over 2}+\gamma+k+l)=(-{\alpha\over 2}+\gamma+k)_l\Gamma(-{\alpha\over 2}+\gamma+k)$, invoke Chu-Vandermonde's theorem and thus write 
$$
\eqalign{\sum_{l=0}^n{{(-n)_l\Gamma(-{\alpha\over 2}+\gamma+k+l)}\over (\gamma)_l\ l!}&=\sum_{l=0}^n{{(-n)_l\ (-{\alpha\over 2}+\gamma+k)_l}\over (\gamma)_l\ l!}\Gamma(-{\alpha\over 2}+\gamma+k)\cr
&=\Gamma(-{\alpha\over 2}+\gamma+k){}_2F_1(-n,-{\alpha\over 2}+\gamma+k;\gamma;1)\cr
&=\Gamma(-{\alpha\over 2}+\gamma+k){({\alpha\over 2}-k)_n\over (\gamma)_n}}\eqno(4.4)
$$
for the finite sum inside the bracket of Eq.(4.3). Consequently, we arrive at 
$$\eqalign{I_{mn}&={1\over 2} 
\beta^{{\alpha\over 2}-\gamma}\sum\limits_{k=0}^m
\Gamma(-{\alpha\over 2}+\gamma+k){({\alpha\over 2}-k)_n\over (\gamma)_n}{(-m)_k\over (\gamma)_k}{1\over k!}
\cr
&={1\over 2}\beta^{{\alpha\over 2}-\gamma}{\Gamma(\gamma)\over (\gamma)_n}
\sum\limits_{k=0}^m(-1)^k{m\choose k}
{\Gamma({\alpha\over 2}+n-k)\Gamma(-{\alpha\over 2}+\gamma+k)\over \Gamma(\gamma+k)\Gamma({\alpha\over 2}-k)},}\eqno(4.5)
$$
wherein we have used the identity ${(-m)_k\over k!}=(-1)^k{m\choose k}$. We further simplify the expression (4.5) by taking note of 
$$
(a-k)_n={(1-a)_k(a)_n\over (1-a-n)_k}
$$ 
in order to justify the relation
$$
\eqalign{I_{mn}&=
{1\over 2} 
\beta^{{\alpha\over 2}-\gamma}
{{\Gamma(\gamma-{\alpha\over 2})({\alpha\over 2})_n}
\over (\gamma)_n}
\sum\limits_{k=0}^m
{(-m)_k(\gamma-{\alpha\over 2})_k(1-{\alpha\over 2})_k
\over (\gamma)_k(1-{\alpha\over 2}-n)_k k!}\cr
&={1\over 2} 
\beta^{{\alpha\over 2}-\gamma}
{{\Gamma(\gamma-{\alpha\over 2})({\alpha\over 2})_n}
\over (\gamma)_n}
{}_3F_2(-m,\gamma-{\alpha\over 2},1-{\alpha\over 2};\gamma,1-{\alpha\over 2}-n
;1),}\eqno(4.6)
$$
which is valid for all values of $\alpha$ and $\gamma$ such that $\gamma-{\alpha\over 2}>0$.
This completes the proof of the lemma\ \P.

It should be noted that Eq.(4.1) is a generalization of Eq.(3.9), and under the limit process of $\alpha \rightarrow 0^+$ Eq.(4.1) reduced to Eq.(3.9). In order to prove this, we proceed by proving
$$
\lim\limits_{\alpha\rightarrow 0^+} ({\alpha\over 2})_n\ {}_3F_2(-m,\gamma-{\alpha\over 2},1-{\alpha\over 2};\gamma,1-{\alpha\over 2}-n
;1)=n!\delta_{mn}.
\eqno(4.7)
$$
This can be demonstrated, by bringing in the series representation of the terminated hypergeometric function ${}_3F_2$, which leads to 
$$
({\alpha\over 2})_n\ {}_3F_2(-m,\gamma-{\alpha\over 2},1-{\alpha\over 2};\gamma,1-{\alpha\over 2}-n
;1)=\sum\limits_{k=0}^m{{(-m)_k(\gamma-{\alpha\over 2})_k(1-{\alpha\over
2})_k(-1)^n\Gamma(1-{\alpha\over 2})}\over k!
(\gamma)_k\Gamma(1-{\alpha\over 2}-n+k)}
$$
by means of the identities
$$(1-{\alpha\over 2}-n)_k={\Gamma(1-{\alpha\over 2}-n+k)\over (1-{\alpha\over
2})_{-n}\Gamma(1-{\alpha\over 2})}=(-1)^n{({\alpha\over 2})_n\Gamma(1-{\alpha\over 2}-n+k)\over
\Gamma(1-{\alpha\over 2})}.$$
Therefore, by taking the limit as $\alpha \rightarrow 0^+$, we can easily see that
$$
\eqalign{\lim_{\alpha\rightarrow 0^+} ({\alpha\over 2})_n\ {}_3F_2(-m,\gamma-{\alpha\over 2},1-{\alpha\over 2};\gamma,1-{\alpha\over 2}-n
;1)&=\sum\limits_{k=0}^m {(-m)_k(-1)^n\over \Gamma(1-n+k)}\cr
&=\sum\limits_{k=0}^m {{(-m)_k(-k)_n}\over k!},}
$$
where we have made use of the relation $(1+k)_{-n}={(-1)^n\over (-k)_n}$. Again, the product ${(-m)_k(-k)_n}$ leads to the fact that the sum
$\sum\limits_{k=0}^m {{(-m)_k(-k)_n}\over k!}$ collapses to $n!$, for the case of $m=k=n$, and 0 otherwise.

The matrix elements of the Hamiltonian (1.3) are now given by means of Lemma (3) in terms of the infinite integral
$$x_{mn}^{-\alpha}=<\psi_m|x^{-\alpha}|\psi_n>=C_mC_n\int\limits_0^\infty x^{2\gamma-\alpha-1} e^{-\beta x^2}{}_1F_1(-n,\gamma,\beta x^2)
{}_1F_1(-m,\gamma,\beta x^2)dx
\eqno(4.8)
$$
Therefore, as consequence of Lemma 2 and Eq.(3.12), the matrix elements now assume the explicit forms
$$
x_{mn}^{-\alpha}=(-1)^{n+m}\beta^{{\alpha\over 2}}{{({\alpha\over 2})_n}\over
(\gamma)_n}{{\Gamma(\gamma-{\alpha\over 2})}\over
\Gamma(\gamma)}\sqrt{{(\gamma)_n(\gamma)_m}\over {n!m!}}{}_3F_{2}(-m,\gamma-{\alpha\over
2},1-{\alpha\over 2};\gamma,1-n-{\alpha\over 2};1);\eqno(4.9)
$$
among which the following is of particular interest
$$
x_{0n}^{-\alpha}=(-1)^{n}\beta^{{\alpha\over 2}}{{({\alpha\over 2})_n}\over
(\gamma)_n}{{\Gamma(\gamma-{\alpha\over 2})}\over
\Gamma(\gamma)}\sqrt{{(\gamma)_n}\over {n!}}.\eqno(4.10)$$

In the case of $\alpha$ being a non-negative even number ($\alpha=2,4,6,\dots$), the hypergeometric function ${}_3F_2$ in Eq.(4.9) may be looked upon as a terminated polynomial of degree $1-{\alpha\over 2}$ instead of an $m$-degree polynomial; thus for $n\geq m$ and $\alpha=2,4,6,\dots$ we have that
$${}_3F_{2}(-({\alpha\over 2}-1), \gamma-{\alpha\over
2},-m;\gamma,1-n-{\alpha\over 2};1)=
\sum\limits_{s=0}^{[{\alpha\over 2}-1]}{{(-m)_s({\gamma-{\alpha\over 2}})_s(1-{\alpha\over 2})_s}\over {s!(\gamma)_s (1-{\alpha\over 2}-n)_s}}.\eqno{(4.11)}
$$ 
As a result hereof, the matrix elements Eq.(4.9) further simplify into the closed form
expressions immediately appearing. These are most suitable for computational purposes as for the case of $\gamma>0$ and $\alpha = 0$, we indeed shall have
$$x_{mn}^{0}=\cases{1& if $n=m$\cr
0& if $n\neq m$\cr}\eqno(4.12)
$$ 
after using Eq.(4.7); as it should have been expected in this specific case.
For the case of $\gamma>1$ and $\alpha=2$, we have from Eq.(4.9) that  
$$
x_{mn}^{-2}=\cases{(-1)^{m+n}
{\beta}{\Gamma(\gamma-1)\over \Gamma(\gamma)}
\sqrt{n!(\gamma)_m\over m!(\gamma)_n}& if $n\geq m$,\cr
\ \cr
(-1)^{m+n}
{\beta}{\Gamma(\gamma-1)\over {\Gamma(\gamma)}}
\sqrt{m!(\gamma)_n\over n!(\gamma)_m}& if $m\geq n$.\cr}
\eqno(4.13)
$$
On the other hand, if the case is $\gamma>2$ and $\alpha=4$, we then have from Eqs.(4.9) and (4.11) that  
$$
x_{mn}^{-4}=\cases{
(-1)^{m+n}\beta^2
{\Gamma(\gamma-2)\over \Gamma(\gamma+1)}
\sqrt{n!(\gamma)_m\over m!(\gamma)_n}[\gamma(n-m+1)+2m]& if $n\geq m$,\cr
\ \cr
(-1)^{m+n}\beta^2
{\Gamma(\gamma-2)\over \Gamma(\gamma+1)}
\sqrt{m!(\gamma)_n\over n!(\gamma)_m}[\gamma(m-n+1)+2n]& if $m\geq n$.\cr}
\eqno(4.14)
$$
As final case that we illustrate, namely $\gamma>3$ and $\alpha=6$, we point to the fact that Eq.(4.9) lets us deduce  
$$
x_{mn}^{-6}=\cases{
(-1)^{m+n}{\beta^3\over 2}
{\Gamma(\gamma-3)\over \Gamma(\gamma+2)}
\sqrt{{n! (\gamma)_m}\over m!(\gamma)_n}\times\cr[(2+n)(1+n)\gamma(\gamma+1)-2m(1+n)
(\gamma-3)(\gamma+1)-m(1-m)(\gamma-2)(\gamma-3)]& if $n\geq m$,\cr
\ \cr
(-1)^{m+n}{\beta^3\over 2}
{\Gamma(\gamma-3)\over \Gamma(\gamma+2)}
\sqrt{m!(\gamma)_n\over n!(\gamma)_m}\times\cr[(2+m)(1+m)\gamma(\gamma+1)-2n(1+m)(\gamma-3)(\gamma+1)-n(1-n)(\gamma-2)(\gamma-3)]& if $m\geq n$.\cr}
\eqno(4.15)
$$
Actually, we can derive similar expressions for all even integers beyond $6$ - i. e. $\alpha=8,10,\dots$. We point to the cases where $m\geq n$, where the derivation is achieved by interchanging the order of the summation of Eq.(3.2) and applying therefater Eq.(4.12) in reversed order. The particular case of $A=0$ (or  $\gamma ={3\over 2}$) and $B=1$, that is allows us, of course, to recover the result of Aguilera-Navarro et al$^{\sref{\agua}}$ as a special case.

\medskip
% -----------------------------------------------------
\noindent{\bf 5. Variational Approach}\medskip
% -----------------------------------------------------
In this section we implement the results developed in the previous section to calculate the matrix elements of $x^{-\alpha}$ by means of a complete orthonormal basis. Thereby we shall be able to introduce a variational treatment of the spiked harmonic oscillator Hamiltonian given by Eq.(1.1). The principle idea is the representation of the Hamiltonian Eq.(1.1) as
$$
H = -{d^2\over dx^2}+ B x^2 +{\lambda\over x^\alpha} = -{d^2\over dx^2}+ B x^2+{A\over x^2}+({\lambda\over x^\alpha}-{A\over x^2});\eqno(5.1)
$$
where at a later point, $A (\neq 0)$ plays the role of an extra degree of freedom, which shall  be determined through a minimization procedure. Although the idea is not complicated, it has many advantages and we mention just a few. First, the range of $\alpha$ is no longer restricted and it can be extended as one pleases, provided the condition $2 \gamma> \alpha$ or more explicitly $A> {1\over 4}(\alpha-2)^2-{1\over 4}$ is satisfied. Second, it substantially reduces the number of basis elements required for the computation of the eigenvalues of the Hamiltonian Eq.(1.1), even for the intermediate region $\lambda\approx 1$. Third, it be can be adapted effectively and allows easy handling by means of symbolic software such as Mathematica. Fourth, the approach of Aguilera-Navarro et al becomes now a special case, namely set $A=0$ and $B=1$. Fifth, it can be easily extended to the case of $N$ dimensions, where the orbital angular momentum number $l$ is arbitrary, with minor modifications only, as we shall see in the next section.  

Let $\psi(x)$ be a {\it trial function} for Hamiltonian $H$ given by Eq.(5.1),
and let us suppose that $\psi(x)$ is expandable as a finite linear combinations of the basis functions $\psi_n(x)$ as given by Eq.(3.13) - i. e.
$$
\psi(x)=\sum\limits_{n=0}^{D-1} a_n \psi_n(x).\eqno(5.2)
$$
The problem now is to minimize the eigenenergies of Eq.(5.2) with respect to the variational parameters $a_n$, $n=0,1,\dots,D-1$ in the finite dimensional subspace $H_{D}$ spanned by the $D$ functions $\psi_0,\psi_1,\dots,\psi_{D-1}$. However, this is equivalent to diagonalizing the Hamiltonian in Eq.(5.2) in the subspace $H_{D}$. By separating the Hamiltonian Eq.(5.2) into two
contributions $H_0=-d^2/dx^2+Bx^2+Ax^{-2}$ and $H_I=\lambda x^{-\alpha}-Ax^{-2}$, we have 
$$
H_{mn}=\int\limits_0^\infty  \psi_m(x) H\psi_n(x)dx\equiv \bra \psi_m|H_0|\psi_n \ket+\bra
\psi_m|H_I|\psi_n\ket,\quad 
m,n=0,1,2,\dots,D-1.\eqno(5.3)
$$
Since the matrix representation of $H_0$ is diagonal in the basis $\{\psi_n\}_0^\infty$, the first term on the right-hand side of Eq.(5.3) yields the exact solution of Gol'dman and Krivchenkov potential Eq.(3.6) - i. e.
$$\bra\psi_m |H_0| \psi_n\ket =2\beta(2n+\gamma)\delta_{mn}\quad\quad(\beta=\sqrt{B},\gamma=1+{1\over 2}\sqrt{1+4A})$$
- and the second term on the right-hand side of Eq.(5.3) is given by
$$\bra
\psi_m|H_I|\psi_n\ket=\lambda \bra
\psi_m|x^{-\alpha}|\psi_n\ket-A\bra
\psi_m|x^{-2}|\psi_n\ket,
$$
where $\bra \psi_m|x^{-\alpha}|\psi_n\ket$ and $\bra \psi_m|x^{-2}|\psi_n\ket$ are given by Eqs. (4.9) and (4.13) respectively.

Two important observations follow from these results. First, the matrix elements of the Hamiltonian (5.1) in terms of the Gol'dman and Krivchenkov basis Eq.(3.13) are given explicitly by ($m,n=0,1,2,..,D-1$, $n\geq m$)
$$\eqalign{
H_{mn}=& 2\beta(2n+\gamma)\delta_{nm}+
(-1)^{n+m}\lambda \beta^{{\alpha\over 2}}{{({\alpha\over 2})_n}\over
(\gamma)_n}{{\Gamma(\gamma-{\alpha\over 2})}\over
\Gamma(\gamma)}\sqrt{{(\gamma)_n(\gamma)_m}\over {n!m!}}\times
\cr 
&{}_3F_{2}(-m,\gamma-{\alpha\over
2},1-{\alpha\over 2};\gamma,1-n-{\alpha\over 2};1)-(-1)^{m+n}
{A\beta\over \gamma-1}
\sqrt{n!(\gamma)_m\over m!(\gamma)_n}\cr}\eqno(5.4)
$$
taken over the $D$-dimensional subspace spanned by basis (3.13). 
The expressions in Eq.(5.4) are highly suitable for the systematic computer-aided calculations of the energy eigenvalues via diagonalization and subsequent minimization of the matrix 
$$\min_{A}\ \hbox{diag}\pmatrix{H_{00}&H_{01}&\dots&H_{0D-1}\cr
		    H_{10}&H_{11}&\dots&H_{1D-1}\cr
		    \dots&\dots&\dots&\dots\cr
		    H_{D-10}&H_{D-11}&\dots&H_{D-1D-1}}.\eqno(5.5)$$  
Secondly, by increasing the matrix dimension $D$, we can always improve these upper energy bounds. In the variational analysis of the ground state energy of the singular potential
$V(x)= x^2+{\lambda x^{-\alpha}}$, Aguilera-Navarro et al$^{\sref{\agua}}$ utilized an orthonormal basis of harmonic oscillator eigenfunctions on the interval $(0,\infty)$ - i.e. the set of
Hermite functions generated by the non-singular harmonic-oscillator potential $x^{2}$. This is equivalent to our basis functions for the case $B =1$ and $A=0$ as mentioned earlier. The shortcomings of their approach, however, are as follows. First, validity only holds for $\alpha <3$. Second, a huge set  of the basis elements is needed to obtain reasonably accurate eigenvalues, and this even for the intermediate region $(\lambda\approx 1)$.  

\medskip
% -----------------------------------------------------
\noindent{\bf 6. The $N$-Dimensional Case}\medskip
% -----------------------------------------------------
In order to extended the scope of our variational analysis to the $N$-dimensional spiked harmonic oscillator Hamiltonian Eq.(1.1), we first determine the exact solutions of the $N$-dimensional Sch\"odinger equation with a Gol'dman and Krivchenkov potential Eq.(3.5). To do this, we notice that the $A$ term of Gol'dman and Krivchenkov potential has the dimensions of kinetic energy, such as the term that appears in higher-dimensional systems. We therefore may replace $A$ in Eq.(3.6) with
$$
A \rightarrow A+\Lambda(\Lambda+1),\quad \Lambda=l+{1\over 2}(N-3)
,\quad N\geq 2,\eqno(6.1)
$$
and obtain thereby an exact solutions of  $N$-dimensional radial Schr\"odinger equation
$$
\bigg(-{d^2\over dx^2}+{\Lambda(\Lambda+1)\over x^2}+
B x^2 +{A\over x^2}\bigg)\psi_{nl}=E_{nl}^N\psi_{nl}.\eqno(6.2)
$$
Such exact solutions are generated from the well-know solutions of harmonic oscillator potential by two simple transformations. We first replace the angular momentum $l$ in the harmonic oscillator energy expression $\beta(4n+2l+3),\ n=0,1,2,\dots$ by $-{1\over 2}+\sqrt{A+(l+{1\over 2})^2}$, and subsequently replace $l$ with $\Lambda$. Thus, the exact eigenvalues of $N$-dimensional Schr\"odinger equation with a Gol'dman and Krivchenkov potential are
$$
E_{nl}^N=2\beta(2n+{\gamma}_N),\quad n,l=0,1,2,\dots,\eqno(6.3)
$$
where $\beta=\sqrt{B}$ and $\gamma_N=1+\sqrt{A+(\Lambda+{1\over 2})^2}$, while the exact eigenfunctions are given by
$$
\psi_{nl}(x)=(-1)^n\sqrt{{2\beta^{\gamma_N} (\gamma_N)_n}\over n! \Gamma(\gamma_N)} x^{\gamma_N-{1\over 2}} e^{-{1\over 
2}\beta x^2}{}_1F_1(-n;\gamma_N;\beta x^2).\eqno(6.4)
$$
The matrix elements in this case turn out to be
$$\eqalign{
\bra \psi_{ml}|x^{-\alpha}| \psi_{nl}\ket=(-1)^{n+m}&\beta^{{\alpha\over 2}}{{({\alpha\over 2})_n}\over
(\gamma)_n}{{\Gamma(\gamma_N-{\alpha\over 2})}\over
\Gamma(\gamma_N)}\sqrt{{(\gamma_N)_n(\gamma_N)_m}\over {n!m!}}\cr
&\times \ {}_3F_{2}(-m,\gamma_N-{\alpha\over
2},1-{\alpha\over 2};\gamma_N,1-n-{\alpha\over 2};1)\quad N\geq 2.\cr}
\eqno(6.5)$$
Matrix elements for the special cases of $\alpha=2,4,6,\dots$ are obtained  by substituting in Eqs.(4.13-15) for $\gamma$ the expression $\gamma_N$, where $\gamma_N = 1+\sqrt{A+(\Lambda+{1\over 2})^2}$. The matrix elements of the Hamiltonian Eq.(5.1) now turn out to be very similar to those in Eq.(5.4), namely
$$H_{mn}= 2\beta(2n+\gamma_N)\delta_{mn}+\lambda\bra\psi_{ml}|x^{-\alpha}|\psi_{nl}\ket -A\bra\psi_{ml}|x^{-2}|\psi_{nl}\ket,\quad N\geq 2.\eqno(6.6)$$
Since the purpose of the present work is to consider the variational analysis of the eigenvalues of the Hamiltonian Eq.(1.1) with Dirichlet boundary condition, we restricted ourself to the  case of $N\geq 2$, in order to avoid problems stemming from the degeneracy of the spectrum in the case of $N=1$. The one-dimensional case Eq.(5.4) on $L_2(0,\infty)$ is recovered by setting $D=3$ and $l=0$. Further, the recovery of the results of Aguilera-Navarro et al is achieved by substituting $A=0$, $B=1$, $D=3$, and $l=0$ in Eq.(6.6). \medskip
% -----------------------------------------------------
\noindent{\bf 7. Numerical results}\medskip
% -----------------------------------------------------
Although the variational method has been used earlier in one form or another, the special form it takes in this present article has a specific purpose. First, the demonstration of the validity of an accurate and uniform approximation not only for the ground state energy of Eq.(1.1), but also for the entire spectrum in arbitrary dimensions with arbitrary angular-momentum number. And second, its applicability to general values of the parameters $(\alpha,\lambda)$ of the potential, subjected to the restriction $\alpha<2\gamma$. The achievement of these purposes is due to the closed forms we have been able to obtain for the matrix-elements Eq.(4.9). At this point we note that the variational results are upper bounds to the energy levels, in accordance with the variational theorem.

For $\alpha<2$ the variational method gives excellent results for arbitrary value of the coupling $\lambda$. The advantage of the minimization process with respect to the parameter $A$ is shown in Table (I). Herein we compare the numerical values obtain by diagonalizationed for $\alpha =0.5$ for different dimensions of the matrix Eq.(5.5) with numerical values obtain by the process of diagonalization and minimization with respect to $A$. There is interesting observation, which must be noted here. For given $\alpha<2$, the dimension of the matrix required for obtaining results of a given accuracy depends on the behavior of the coupling $\lambda$. For $\lambda$ small however, we can get away with a  matrix Eq.(5.5) of smaller dimension as compared with case where $\lambda$ is large. To further illustrate this point, a one-dimension subspace ($1\times 1$ matrix) is adequate for calculating the eigenvalue $3.001~128$ (exact to the accuracy quoted) for $\lambda = 0.001$ when $\alpha=1$; but for $\lambda =10$ on the other hand, a matrix at least of size $80\times 80$ is necessary to obtain the eigenvalue $10.577~48$. In the case of $\alpha =1 $, we present in Table (II) a comparison of our numerical results to some previous work$^{\sref{\aguc}}$. For $\alpha>2$, the situation is completely reversed.    

For $\alpha=2$, the first variational approximation (subspace of dimension one) to the ground state eigenvalues of the spiked harmonic oscillator Hamiltonian is
$$E_0(\alpha=2)=2\beta\gamma+{\lambda \beta\over \gamma -1}-{A\beta\over \gamma-1},\quad \gamma=1+{1\over 2}\sqrt{1+4A}.\eqno(7.1)
$$
The minimization of this expression with respect to the parameter $A$ can be easily performed and leads to $A=\lambda$. On substituting this back into Eq.(7.1) we get $E_0=2\beta (1+{1\over 2}\sqrt{1+4A})$, which is the exact result quoted in Eq.(3.6) for $n=0$.

For arbitrary $\alpha$, the first variational approximation (subspace of dimension 1) of the ground state eigenvalues of the Hamiltonian Eq.(1.1) is
$$
\epsilon_0=\min\limits_A E_0=\min\limits_A\{2\beta\gamma+\lambda \beta^{\alpha\over
2}{{\Gamma(\gamma-{\alpha\over 2})}\over {\Gamma(\gamma)}}-{A\beta\over \gamma -1}\},\quad
\gamma=1+{1\over 2}\sqrt{1+4A}.\eqno(7.2)
$$
When $D=2$, i.e. subspace of dimension 2, the diagonalization can also be performed analytically via the secular equation, that is to say by means of the expression
$$
\epsilon_\pm=\min\limits_A E_\pm=\min\limits_A\{{1\over 2}[(H_{00}+H_{11})\pm
\sqrt{(H_{00}-H_{11})^2+(2H_{01})^2}]\}.\eqno(7.3)
$$
where $\epsilon_0=\epsilon_{-}$ and $\epsilon_1=\epsilon_{+}$. Moreover, one can obtain analytic expressions for upper bound eigenvalues$^{\sref{\halc}}$ in this case.
 
Because of the simple formulas for the matrix elements in the cases of $\alpha=4$ and $\alpha=6$  given by (4.14) and (4.15) respectively, the determination of the energy values to any desired accuracy has now been reduced to an easy task. A heuristic scheme for ascertaining the eigenvalues to any required number of digits is as follows. The eigenvalues obtained from successive levels, such as ($1\times 1, 2\times 2, \dots)$, of the truncated matrix (5.5) are compared, and the calculation ceases when the successive eigenvalue agree with each other up to the prescribed decimal place. It's sufficient, therefore, to use the $N$-dimensional case for the matrix elements Eq.(6.5) for the calculation purpose. To recover the 1-dimensional case, we may set $D=3$ and $l=0$, and to cover Aguilera-Navarro et al's results, we set $A=0, B=1$, $D=3$, and $l=0$. Table (III)  
illustrate the use of this procedure for the case of $\alpha=4$ and $\lambda=1000$ in the dimensions $N=2$ to $10$.
 
Another advantage of the variational approach presented herein is the amount of information that we get about the spectrum of the Hamiltonian Eq.(1.1) every time we compute the eigenvalues via the diagonalization and minimization. Indeed, we obtain for an arbitrary matrix (5.5) of size $D$ a set of upper bounds for the eigenvalues $E_0,E_1,..., E_{D-1}$. Each can be improved by either an increase in the dimension of the matrix, or by extracting the desired level through the diagonalization and thereafter minimizing with respect to parameter $A$, which is illustrated in Table(IV). 

For $\alpha>2$ and small values of $\lambda$, the variational method is still applicable, however the eigenvalues converges very slowly, and an immense number of matrix elements are needed in order to obtain accurate results. In principle, the method is still effective as we see from Table (IV), but at the expense of using a huge number of matrix elements. On the other hand, we obtained accurate eigenvalues with the use of a modest number of matrix elements for $\lambda$ varying between $1000$ to $0.01$ as Table (V) indicates. An interesting explanation of this lies in the following comment. Let us look at the perturbed Hamiltonian 
$$
H = -{d^2\over dx^2}+ x^2+{A\over x^2}+\epsilon({\lambda\over x^\alpha}-{A\over x^2})\eqno(7.4)
$$
instead of the one given by Eq.(5.1), and solve the matrix eigenvalue problem by expanding the determinant in powers of $\epsilon$ up to and including $\epsilon^2$. It is well known$^{\sref{\agua}}$ that we shall end up with the perturbation-like formula (after setting $\epsilon=1$)
$$
E=E_0+\lambda \bra \psi_0|x^{-\alpha}|\psi_0\ket-A\bra \psi_0|x^{-2}|\psi_0\ket-\sum_{n\neq 0}^D{|\bra \psi_0|\lambda x^{-\alpha}-Ax^{-2}|\psi_n\ket|^2\over E_n-E_0},\eqno(7.5)
$$
where the summation on the right-hand side is finite and $D$ is the number of the basis functions used. By analyzing this sum, we shall come to understand the slow convergence of the eigenvalues calculated by means of the variational approach. For the sake of simplicity, let us restrict ourselves to the case of $\alpha=4$. With the aid of Eq.(4.13) and Eq.(4.14), the sum in Eq.(7.5) becomes $$
\eqalign{
\sum_{n\neq 0}^D{|\bra \psi_0|\lambda x^{-4}-Ax^{-2}|\psi_n\ket|^2\over E_n-E_0}
&=
{1\over 4}{\Gamma^2(\gamma-2)\over \Gamma^2(\gamma)}
\bigg[\lambda^2\sum\limits_{n=1}^D {(n+1)!\over (\gamma)_n}+2\lambda(\lambda-A(\gamma-2))\sum\limits_{n=1}^D {(n+1)!\over n(\gamma)_n}\cr
&+(\lambda-A(\gamma-2))^2\sum\limits_{n=1}^D {(n+1)!\over n^2(\gamma)_n}\bigg]
\cr}\eqno(7.6)
$$ 
in this case. For $D\rightarrow \infty$, the sums on the right hand side of Eq.(7.6) have closed form expressions in terms of hypergeometric functions, particularly 
$$
\eqalign{
\sum_{n\neq 0}^\infty&{|\bra \psi_0|\lambda x^{-4}-Ax^{-2}|\psi_n\ket|^2\over E_n-E_0}
=
{1\over 2\gamma}{\Gamma^2(\gamma-2)\over \Gamma^2(\gamma)}
\bigg[\lambda^2{}_2F_1(3,1;\gamma+1;1)+2\lambda(\lambda-A(\gamma-2))\cr&\times {}_3F_2(3,1,1;2,\gamma+1;1)+(\lambda-A(\gamma-2))^2{}_4F_3(3,1,1,1;2,2,\gamma+1;1)\bigg].
\cr}\eqno(7.7)
$$ 
Herein the conditions of the convergence of hypergeometric functions $^{\sref{\slat}}$ guarantee that the function ${}_2F_1$ converges for $\gamma>3$, the function ${}_3F_2$ converges for $\gamma>2$ and ${}_4F_3$ converges for $\gamma>1$. Thus, Eq.(7.7) holds for $\gamma>3$ or $A>3.75$ in general, but for $\alpha=4$, the matrix elements Eq.(4.9) holds for $\gamma>2$ or $A>0.75$. This demonstrates the difficulties one encounters with the variational method for small value of $\lambda$. To understand this further, we note the relation between $\lambda$ and the parameter $A$. Thus for $\lambda$ small, $A$ is small; on the other hand, $\lambda$  large implies $A$ large. This is evident from the first variational approximation, namely Eq.(7.2) yields
$$
\epsilon_0=\min\limits_A\{2\gamma+{\lambda \over {(\gamma-1)(\gamma-2)}}-{A\over \gamma -1}\},\quad
\gamma=1+{1\over 2}\sqrt{1+4A}.\eqno(7.8)
$$
After differentiating Eq.(7.8) with respect to $A$, it can shown that the relation between $A$ and $\lambda$ is implicitly given by the relation
$$
\lambda={{(\gamma-2)^2(4(\gamma-1)^2-1)}\over 4(2\gamma-3)},\eqno(7.9)
$$
from which it is quite easy to see that $\lambda$ is an increasing function of $\gamma$ for all $\gamma>2$. Thus for Eq.(7.7) to converge, we must have $\gamma>3$ - i.e. $\lambda>1.25$. This demonstrates that for $\lambda<1.25$, the accuracy of the eigenvalue calculations by means of the variational method, necessitates a matrix of very large order. 
\medskip
% -----------------------------------------------------
\noindent{\bf 8. Conclusion}\medskip
% -----------------------------------------------------
In this paper we have carried our study of the spiked harmonic-oscillator problem further by expressing the Hamiltonian $H$ given by Eq.(1.1) as the perturbation of the singular Gol'dman and Krivchenkov Hamiltonian $H_0 = -{d^2\over dx^2}+ B x^2+{A\over x^2}$, where the expression $\lambda x^{-\alpha} - A x^{-2}$ is looked upon as the perturbation term. 
We have provided a compact proof of the fact that the eigenfunctions of order $0$ generated by $H_0$ form a suitable singularity-adapted basis for the appropriate Hilbert space of the full problem. Our principal results are fourfold: (1) the proof of the completeness of the orthogonal set of normalized eigenvalues of $H_0$ by means of the domains of convegence, absolute convergence, holomorphy of the Laplace Transform; (2) the derivation of the compact closed form (1.3) for the matrix elements $x^{-\alpha}_{mn}$ via infinite integrals of products of two confluent hypergeometric functions; (3) a variational method applied to our pertubation of the Hamiltonian $H_0$, which proves to be very effective for accurate calculation the eigenvalues of the spiked harmonic oscillator, as the accompanying numerical calculations verify; (4) we show that this variational approach is applicable to the entire discrete spectrum for the spiked harmonic oscillator in $N$-dimensions.   
\bigskip
\noindent{\bf Acknowledgment}
\medskip Partial financial support of this work under Grant No. GP3438
from the 
Natural Sciences and Engineering Research Council of Canada is gratefully 
acknowledged by one of us [RLH].
\np
%------------------------------------- 
\references{1}
%-------------------------------------
\vfil\eject
% -------------------------------------------------------------------% --------------------------------------------------------------------------
\noindent {\bf Table (I)}~~~The rate of convergency for the upper bound of the ground state energy $E_0$ of the Schr\"odinger equation 
$(-\Delta+x^2+{\lambda\over x^{0.5}})\psi=E_0\psi$ for $\lambda=0.1$ and $\lambda = 1000$. $E_0^{A=0}$
obtained by diagonalization of the $D\times D$ matrix elements, $E_0^{A}$  obtained by diagonalization of the $D\times D$ matrix elements then minimizing with respect to the parameter $A$. 
\bigskip

\noindent\hfil\vbox{%
\offinterlineskip
\tabskip=0pt
\halign{\tabskip=5pt
\vrule#\strut& #\strut\hfil&\vrule#\strut&\hfil#\strut\hfil&\vrule#\strut&
\hfil#\strut\hfil&\vrule#\strut&\hfil#\strut\hfil&\vrule#\strut&\hfil#\strut\hfil&
\vrule#\strut
\tabskip=0pt\cr
\multispan2&\multispan9\hrulefill\cr
\multispan2&\multispan4\vrule\hfil$\lambda=0.1$\hfil&&\multispan3\hfil $\lambda=1000$\hfil&\cr
\multispan2&\multispan5&\omit&\omit
\vrule\cr\noalign{\hrule}
\multispan2\vrule ${\bf D\times D}$&& $E_0^{A=0}$&&
		$E_0^{A}$&& $E_0^{A=0}$ && $E_0^{A}$ &\cr
\noalign{\hrule}
&$1\times 1$&&3.102~277&&3.102~185&&1025.765~672 &&415.934~312&\cr
\noalign{\hrule}
&$2\times 2$&&3.102~167&&3.102~149&&746.081~846&&415.932~051&\cr
\noalign{\hrule}
&$3\times 3$&&3.102~151&&3.102~143&&642.417~430&&415.890~659&\cr
\noalign{\hrule}
&$5\times 5$&&3.102~143&&3.102~141&&549.825~333&&415.889~798&\cr
\noalign{\hrule}
&$10\times 10$&&3.102~140&&3.102~139&&461.349~666&&415.889~785&\cr
&\ &&\ &&(Exact) &&\ &&(Exact)&\cr
\noalign{\hrule}
}
}

\vfil\eject

\noindent {\bf Table (II)}~~~A comparison between the results $E^{RP}$ by means of Ricccati-Pad\'e$^{\sref{\aguc}}$, and $E^{R}$ by means of renomalized series$^{\sref{\aguc}}$ and the result of the present work $E$ (correct to 7 digit shown) for $\alpha=1$ and various values of the coupling $\lambda$.

\bigskip

\hskip 1 true in
\vbox{\tabskip=0pt\offinterlineskip
\def\tablerule{\noalign{\hrule}}
\def\vr{\vrule height 12pt}
\halign to300pt{\strut#\vr&#
\tabskip=1em plus2em
&\hfil#\hfil
&\vrule#
&\hfil#\hfil
&\vrule#
&\hfil#\hfil
&\vrule#
&\hfil#\hfil
&\vr#\tabskip=0pt\cr
\tablerule&&$\lambda$&&$E^{RP}$&&$E^{R}$&&$E$&\cr\tablerule
&&0.001&&$3.001~128$&&$3.001~143$&&$3.001~128$&\cr\tablerule
&&0.01&&$3.011~276$&&$3.011~417$&&$3.011~276$&\cr\tablerule
&&0.1&&$3.110~9$&&$3.113~386$&&$3.112~068$&\cr\tablerule
&&1&&$4.057~906$&&$4.064~649$&&$4.057~888$&\cr\tablerule
&&10&&$10.577~483$&&$10.577~825$&&$10.577~485$&\cr\tablerule
}}

\vfil\eject
\noindent {\bf Table (III)}~~~Upper bounds $E^U$ for 
$H=-\Delta+x^2+{1000\over x^{4}}$ for dimension $N=2$ to 10, 
obtained by diagonalization then minimization of the $30\times 30$ matrix elements. The ``exact'' values $E$ were obtained by 
direct numerical integration of Schr\"odinger's equation.
\bigskip 

\hskip 1 true in
\vbox{\tabskip=0pt\offinterlineskip
\def\tablerule{\noalign{\hrule}}
\def\vr{\vrule height 12pt}
\halign to200pt{\strut#\vr&#
\tabskip=1em plus2em
&\hfil#\hfil
&\vrule#
&\hfil#\hfil
&\vrule#
&\hfil#\hfil
&\vr#\tabskip=0pt\cr
\tablerule&&$N$&&\bf $E^U$&&\bf $E$ &\cr\tablerule
&&2&&$21.350~246$&&$21.350~246$&\cr\tablerule
&&3&&$21.369~463$&&$21.369~463$&\cr\tablerule
&&4&&$21.427~056$&&$21.427~056$&\cr\tablerule
&&5&&$21.522~859$&&$21.522~859$&\cr\tablerule
&&6&&$21.656~596$&&$21.656~596$&\cr\tablerule
&&7&&$21.827~883$&&$21.827~883$&\cr\tablerule
&&8&&$22.036~232$&&$22.036~232$&\cr\tablerule
&&9&&$22.281~057$&&$22.281~057$&\cr\tablerule
&&10&&$22.561~680$&&$22.561~680$&\cr\tablerule
}}

\vfil\eject

\noindent {\bf Table (IV)}~~~For arbitrary size $D\times D$ of the matrix (5.5) for 
$H=-\Delta+x^2+{1000\over x^{4}}$, a set of upper bounds for the eigenvalues $E_0,E_1,..., E_{D-1}$ that can improve by increasing the size $D$.
\bigskip
$$
\vbox{\offinterlineskip\cleartabs
\def\hr{\vrule height .4pt width 5em}
\def\vr{\vrule height12pt depth 5pt}

\+ \hr&\hr&\hr&\hr&\hr&\hr&\hr&\cr
\+ \vr\hfil$1\times 1$&\vr\hfil$2\times 2$&\vr\hfil$3\times 3$&\vr\hfil$4\times 4$&\vr\hfil$5\times 5$&\vr\hfil$6\times 6$&\vr\hfil$7\times 7$&\vr\cr
\+ \hr&\hr&\hr&\hr&\hr&\hr&\hr&\cr
\+ \vr\hfil$21.427~79$&\vr\hfil$21.382~12$&\vr\hfil$21.374~00$&\vr\hfil$21.370~07$&\vr\hfil$21.369~72$&\vr\hfil$21.369~51$&\vr\hfil$21.369~46$&\vr\cr
\+ \hr&\hr&\hr&\hr&\hr&\hr&\hr&\cr
\+ &\vr\hfil$26.298~42$&\vr\hfil$26.189~48$&\vr\hfil$26.166~99$&\vr\hfil$26.155~44$&\vr\hfil$26.154~18$&\vr\hfil$26.153~40$&\vr\cr
\+ &\hr&\hr&\hr&\hr&\hr&\hr&\cr
\+ &\ &\vr\hfil$31.097~17$&\vr\hfil$30.919~24$&\vr\hfil$30.878~34$&\vr\hfil$30.856~56$&\vr\hfil$30.851~94$&\vr\cr
\+ &\ &\hr&\hr&\hr&\hr&\hr&\cr
\+ &\ &\ &\vr\hfil$35.834~86$&\vr\hfil$35.587~50$&\vr\hfil$35.525~79$&\vr\hfil$35.492~11$&\vr\cr
\+ &\ &\ &\hr&\hr&\hr&\hr&\cr
\+ &\ &\ &\ &\vr\hfil$40.520~33$&\vr\hfil$40.205~49$&\vr\hfil$40.121~62$&\vr\cr
\+ &\ &\ &\ &\hr&\hr&\hr&\cr
\+ &\ &\ &\ \ &\ &\vr\hfil$45.160~79$&\vr\hfil$44.781~42$&\vr\cr
\+ &\ &\ &\ &\ &\hr&\hr&\cr
\+ &\ &\ &\ \ &\ &\ &\vr\hfil$49.762~16$&\vr\cr
\+ &\ &\ &\ &\ &\ &\hr&\cr
}
$$

\vfil\eject
\noindent {\bf Table (V)}~~~Upper bounds $E^U$ for 
$H=-{d^2\over dx^2}+x^2+{\lambda\over x^{\alpha}}$\ ($\alpha =4$ and $6$), for small  values of $\lambda$, 
obtained by diagonalization of the $D\times D$ matrix elements then minimizing with repect to the parameter $A$. 
\bigskip
\noindent\hfil\vbox{%
\offinterlineskip
\tabskip=0pt
\halign{\tabskip=5pt
\vrule#\strut&#\strut\hfil&\vrule#\strut&\hfil#\strut\hfil&\vrule#\strut&\hfil#\strut\hfil&\vrule#\strut&\hfil#\strut\hfil&\vrule#\strut&\hfil#\strut\hfil&\vrule#\strut&\hfil#\strut\hfil&\vrule#\strut&\hfil#\strut\hfil&\vrule#\strut\tabskip=0pt\cr
\multispan2&\multispan4{\hrulefill}&\multispan4{\hrulefill}&\multispan4\hrulefill\cr
\multispan2&\multispan4\vrule\hfil$\lambda=0.0025$\hfil&&\multispan3\hfil$\lambda=0.005$\hfil&&\multispan3\hfil $\lambda=0.01$\hfil&\cr
\multispan5&\multispan5&\omit&\omit\vrule\cr\noalign{\hrule}
\multispan2\vrule~E[Ref.]&&$\alpha=4$&&
		$\alpha=6$&&$\alpha=4$&&$\alpha=6$&&$\alpha=4$&&$\alpha=6$&\cr
\noalign{\hrule}
&$E^{DK}$&&3.106~70&&3.353~95&&3.148~39&&3.423~02&&3.205~27&&3.505~74&\cr
\noalign{\hrule}
&$E^H$&&3.037~61&&3.068~22&&3.053~19&&3.081~13&&3.075~22&&3.096~48&\cr
\noalign{\hrule}
&$E^K$&&3.106~81&&3.353~92&&3.148~35&&3.422~88&&3.205~07&&3.505~45&\cr
\noalign{\hrule}
&$E^P$&&3.103~77&&3.343~05&&3.146~64&&3~413~16&&3.204~42&&3.496~88&\cr
\noalign{\hrule}
&$E^L$&&3.106~81&&3.353~92&&3.148~35&&3.422~88&&3.205~07&&3.505~45&\cr
\noalign{\hrule}
&$E^U$&&3.107~95&&3.354~095&&3.149~00&&3.422~95&&3.205~48&&3.505~49&\cr
\noalign{\hrule}}
}
\item{$^{DK}$}{From Detwiler and Klauder (1975).}
\item{$\ \ ^{H}$}{From Harrell (1977).}
\item{$\ \ ^{K}$}{From Killingbeck (1982). Richardson extrapolation. Correct to the 6 digits shown.}
\item{$\ \ ^{P}$}{From Pad\'e approximant technique (W. Solano-Torres et al, 1992).}
\item{$\ \ ^{L}$}{From Lanczos/grid Method (W. Solano-Torres et al, 1992). Correct to the 6 digits shown.}
\item{$\ \ ^{E^U}$}{From the present work.}

\vfil\eject

\noindent {\bf Table (VI)}~~~A comparison between the results $E^F$ of Fern\'andez$^{\sref{\fer}}$, and the results $E^{HS}$ of Hall et al$^{\sref{\hnas}}$ and the result of the present work $E^{HSK}$ (correct to 7 digit shown) for $\alpha=4$ and $\alpha=6$ and various values of the coupling $\lambda$.
\bigskip

\noindent\hfil\vbox{%
\offinterlineskip
\tabskip=0pt
\halign{\tabskip=5pt
\vrule#\strut&\hfil#\strut\hfil&\vrule#\strut&\hfil#\strut\hfil&\vrule#\strut&\hfil#\strut\hfil&\vrule#\strut&\hfil#\strut\hfil&\vrule#\strut&\hfil#\strut\hfil&\vrule#\strut&\hfil#\strut\hfil&\vrule#\strut&\hfil#\strut\hfil&\vrule#\strut
\tabskip=0pt\cr
\multispan2&\multispan7\hrulefill&\multispan6\hrulefill\cr
\multispan2&\multispan6\vrule\hfil$\alpha=4$\hfil&&\multispan5\hfil$\alpha=6$\hfil&\cr
\noalign{\hrule}
\multispan2\vrule\ $\lambda$\ && $E_0^{F}$&&$E_0^{HS}$&& $E_0^{HSK}$&&$E_0^{F}$&& $E_0^{HS}$&&$E_0^{HSK}$ &\cr
\noalign{\hrule}
&1000&&21.384~46&&21.370~26&&21.369~462&&12.737~60&&12.725~65&&12.718~617&\cr
\noalign{\hrule}
&100&&11.292~41&&11.265~86&&11.265~080&&8.422~60&&8.420~96&&8.413~358&\cr
\noalign{\hrule}
&10&&6.649~78&&6.609~66&&6.606~622&&6.016~40&&6.014~94&&6.003~209&\cr
\noalign{\hrule}
&1&&4.548~79&&4.504~16&&4.494~179&&4.676~88&&4.684~97&&4.659~940&\cr
\noalign{\hrule}
&0.1&&3.626~44&&3.600~44&&3.575~557&&4.019~15&&4.042~84&&3.915~665&\cr
\noalign{\hrule}
&0.01&&3.237~75&&3.249~80&&3.205~486&&3.524~93&&3.580~70&&3.505~492&\cr
\noalign{\hrule}
}
}
\vfil\eject

\end